\documentclass[twocolumn,showpacs,prl]{revtex4} 

\usepackage{latexsym}
\usepackage{dcolumn}
\usepackage{bm}
\usepackage{latexsym,epsfig,amssymb,amsmath}
\usepackage{color}

\definecolor{darkblue}{rgb}{0,0,0.5}
\definecolor{darkgreen}{rgb}{0,0.4,0}

\usepackage{hyperref}
\usepackage{bm}

\definecolor{orange}{rgb}{1,0.5,0}
\usepackage{color}

\usepackage{dsfont}

\renewcommand{\vec}{\bm}

\newcommand{\karman}{von ~ K\'arm\'an~}
\newcommand{\degree}{\ensuremath{^\circ}}

\newcommand{\pa}{^\mathbb{P}} 
\newcommand{\ie}{\emph{i.e.~}} 
\newcommand{\fig}[1]{Fig.~\ref{#1}} 

\hypersetup{pdftex=true, colorlinks=true, breaklinks=true, linkcolor=darkblue, menucolor=darkblue,  urlcolor=darkblue}

\begin{document}

\title{Rotational intermittency and turbulence induced lift  \\ {experienced by
large} particles in a turbulent flow.}

\author{Robert Zimmermann,  Yoann Gasteuil, Mickael Bourgoin, Romain Volk, Alain Pumir,  Jean-Fran\c{c}ois Pinton}

\affiliation{Laboratoire de Physique de l'\'Ecole Normale 
Sup\'erieure de Lyon, UMR5672, CNRS \& Universit\'e de Lyon, 46 All\'ee d'Italie, 69007 Lyon, France}

\begin{abstract}
The motion 
of a large, neutrally buoyant, particle, freely advected by a turbulent flow is determined experimentally.
We demonstrate that both the translational and angular accelerations
exhibit very wide probability distributions, a manifestation of intermittency.
The orientation of the angular velocity with respect to the trajectory,
as well as the translational acceleration conditioned on the spinning velocity 
provide evidence of a lift force acting on the particle.
\\
\end{abstract}
\maketitle

\begin{figure}[h!]
\centerline{\includegraphics[width=6cm]{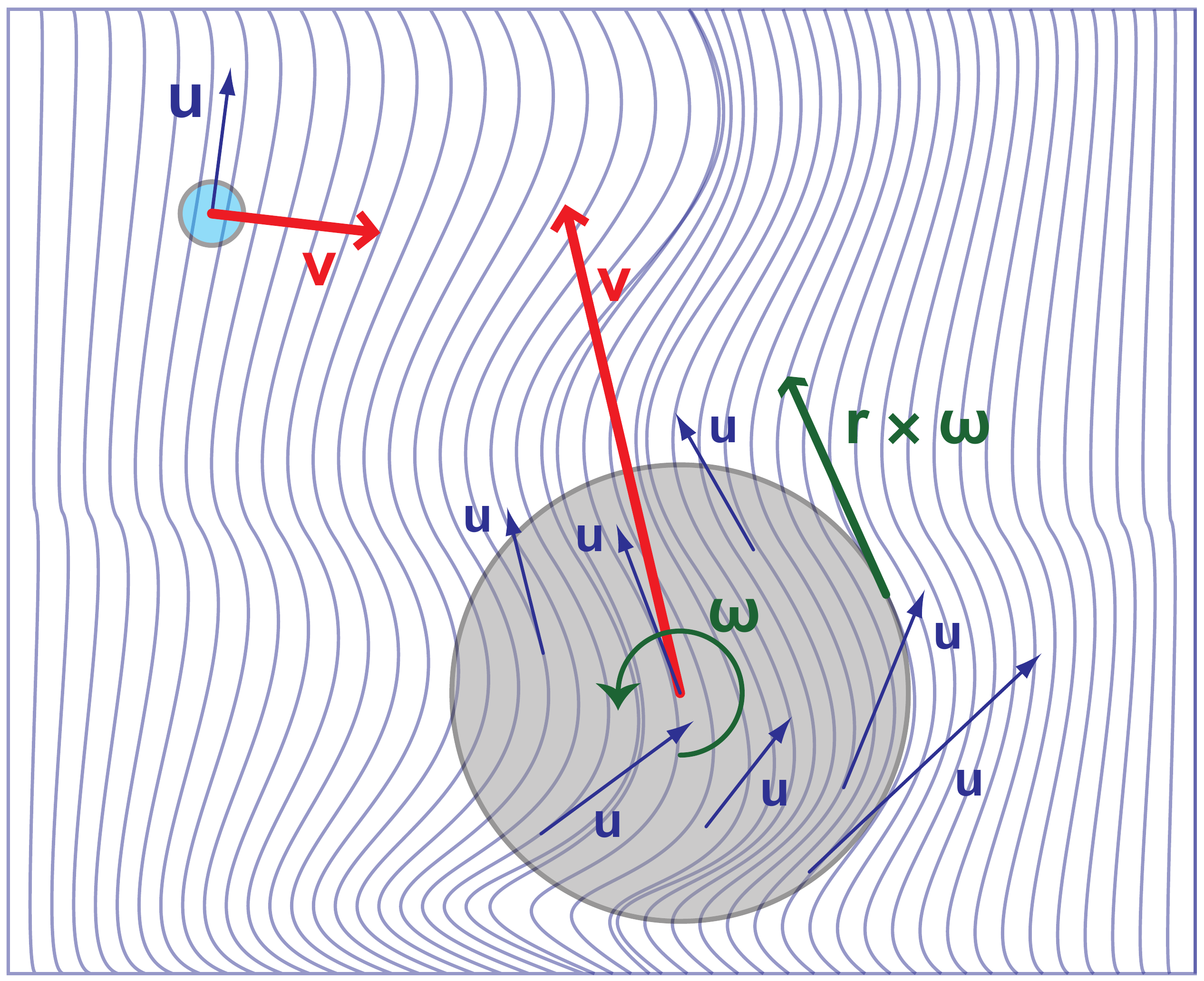}}
\caption{Particles in a turbulent flow. Whereas the flow around the small 
particle is smooth, it exhibits significant spatial variations around the large particle. }
\label{flow_sketch} 
\end{figure}

The description of a solid object freely advected in a fluid requires, in addition to its translational degrees of freedom characterizing its position, $3$ rotational degrees of freedom, specifying its orientation  with respect to a reference frame. The evolution of its position and of its orientation depends, according to Newton's laws, on the forces and torques acting on the particle at each instant, which result from the interaction between the object and the turbulent flow. 
Their determination raises 
challenging issues. The problem has been solved to a large extent for spherical particles of size $D$ much smaller than the smallest length scale of the flow, the Kolmogorov scale~$\eta$~\cite{Gatignol:1983kl,Maxey:1983kl}. Because of the small size of the particle, the flow around it is locally laminar -- see \fig{flow_sketch} -- therefore the equation governing the particle velocity, $\vec{v}$, can be determined by solving the fluid (Stokes) equations once the fluid velocity, $\vec u$, is known. In the simplest case, the particles are subject to the Stokes drag and the added mass term~\cite{Elghobashi:1992dp}, thus the velocity $\vec{v}$ can be determined by solving a simple differential equation. 
For $D \rightarrow 0$, the velocity of a neutrally buoyant particle reduces to the fluid velocity, $\vec{u}$, so the particle behaves as a fluid tracer. 
This property is crucial for several experimental techniques
~\cite{Tropea:2007eu}. 
Interestingly, the translational and rotational degrees of freedom completely decouple in the limit $D \rightarrow 0$~\cite{Gatignol:1983kl,Maxey:1983kl}.

As suggested by \fig{flow_sketch}, the case of particles of radius larger than $\eta$ is conceptually much more difficult. Experiments have shown that upon increasing the ratio $D/\eta$ from $1$ to $40$, the variance of the particle acceleration (\ie of the forces) decreases as $(D/\eta)^{-2/3}$~\cite{Qureshi:2007pb,Voth:2002hc}. The fluctuations of force are still non--Gaussian up to $D/\eta \lesssim 40$~\cite{Qureshi:2008qe,Brown:2009fu,Volk:2010oz}. A full derivation of the equation of motion of a large particle is still not available. 
The possible existence of a lift force when the 
particle Reynolds number $Re_p \equiv u_{rms} D/\nu$ is large,
coupling rotational and translational dynamics, has been the subject of several studies~\cite{Auton:1988tg,Loth:2009oq}. It can be viewed as a generalization of the Magnus force, derived in an inviscid, laminar flow: 
${\vec{F}}_{\rm lift} = C_{\rm lift}\,\vec\omega\pa \times {\vec{v}}_{\rm rel}$, where 
$\vec\omega\pa$ is the rotation of the particle in the flow.
A lift has been measured in laboratory experiments, when the flow is steady and laminar~\cite{Nierop:2007sp,Rastello:2009ij}. 
The flow conditions in these experiments are clearly very different from
the case of a particle in turbulence, schematically
represented in \fig{flow_sketch}. The very existence of any lift force 
in these conditions is thus not obvious.

\begin{figure*}[t]
\centering
\includegraphics[width=0.85\textwidth]{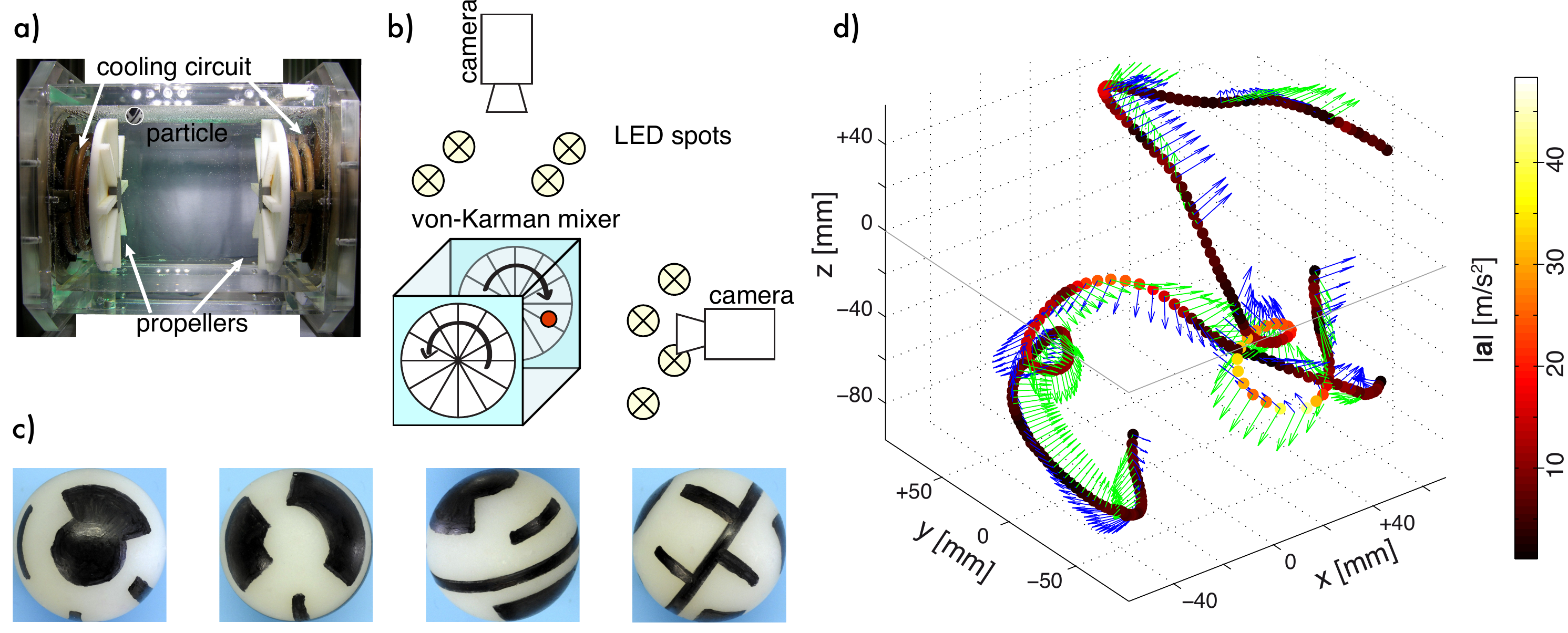}
\caption{(a) The flow domain in between the impeller has characteristic lengths $H = 2R = 20$~cm; the disks have radius $R=10$~cm, fitted with straight blades $1$~cm in height; (b) It is illuminated by high power LEDs and sequences of 8~bit gray scale images are recorded using high-speed cameras; (c) an adequate texture painted on the particle allows the tracking of its orientation; (d) Example of particle tracks and orientations (the green and blue arrows mark North-South and East-West directions).}
\label{setup} 
\end{figure*}

Here, we measure simultaneously the translational and rotational motion of a neutrally buoyant spherical particle, whose diameter is a fraction of the integral size of the turbulence, and report the evidence  of a lift force. We first briefly  describe our 6-dimensional tracking technique~\cite{Gasteuil:2009la,Ye:1992rw,Zimmermann:2010qa},  then we present our results concerning the  intermittency of the translational \emph{and} rotational velocity and acceleration. Last, we discuss how the translational and rotational degrees of freedom of the particle couple, resulting in a lift force. \\

The flow and tracking technique are described in details 
in~\cite{Zimmermann:2010qa}. A \karman (VK) flow, as widely used for turbulence
studies~\cite{Monchaux:2006dz,Ouellette:2006fy,Voth:2002hc}, is generated in 
the gap between two counter-rotating impellers, see \fig{setup}(a). 
In order to be able to perform direct optical measurements, the container is built with flat Plexiglas side walls, so that the cross section of the vessel has a
square shape. 

In the results reported here, the driving disks are rotated at $3$~Hz, which corresponds to a power input in the system of the order of $\varepsilon \sim 1.7$~W/kg. The integral time and length scales are $L_{\text{int}} = 3$~cm,  $T_L = 0.3$~s, so that the dissipative space and time scales are respectively $\eta \sim 30\,\mu$m and $\tau_\eta \sim 1$~ms. The flow Reynolds number based on the Taylor micro-scale is $R_\lambda \sim 300$.  

A PolyAmide sphere with density $\rho_p = 1.14 \; {\rm g}.{\rm cm^{-3}}$ and diameter $D=18$~mm is made neutrally buoyant by adjusting the density of the fluid by addition of glycerol to water (the final density mismatch is less than $\Delta \rho \,/ \rho = 10^{-4}$). The spheres are homogeneous, \ie the particle center of mass coincides with its geometrical center. Their size is comparable to the  integral size of the flow, corresponding to $D/\eta \sim 600$ and $D/L_{\text{int}} \sim 0.6$. Its motion is recorded at a frame-rate equal to $600$~Hz, using 2 high-speed video 
cameras (Phantom V12, Vision Research Inc.) positioned at 90 degrees. The measurement volume lies within 75\% of the radial and axial distances on either side of the flow center, \ie extends to regions where anisotropy and inhomogeneity are known to play a role. The position of the sphere is determined using a position tracking algorithms, as in~\cite{Voth:2002hc}. The time-resolved determination of the rotational degrees of freedom~\cite{Gasteuil:2009la,Zimmermann:2010qa} is carried out by painting a pattern that enables a determination of the particle's orientation from the camera images, cf.  \fig{setup}(c). The orientation tracking algorithm then: (i) compares the sphere's picture with synthetic images and identifies a set of possible orientations; (ii) from the set of possible candidates at successive instants, a \emph{Flow} algorithm identifies a likely time series; (iii) a post-treatment adjusts remaining ambiguities, using the 2 independent views from the 2 cameras. After processing, all trajectories with a duration longer than $0.25~T_L$ are analyzed -- corresponding to  $3434$ trajectories,  with duration ranging between $0.25$ and $3~T_L$. A few particular examples are shown in \fig{setup}(c), showing both the position, color coded for the amplitude of the acceleration, together with two unit vectors 
fixed in the reference frame of the particle.\\

We find that the behavior of the translational degrees of freedom is qualitatively very similar to the results obtained for much smaller particles~\cite{Brown:2009fu,Qureshi:2007pb,Qureshi:2008qe,Volk:2010oz} -- the size of the particles studied so far did not exceed $ D \lesssim 40 \eta$, whereas ours has $D \approx 600 \eta$. The particle velocity has a quasi-Gaussian distribution; each component has zero mean and an $rms$ intensity of fluctuations of about $60$~cm/s, for impellers rotating at 3~Hz. This is of the order of 30\% of the impeller tip speed, and also of  the order of the $rms$ velocity fluctuations of tracers in the same flow. This leads to the consistent notion that the velocities of the particle are of the order of the large scale swirls generated in the VK setup.
The particle Reynolds number $Re_p \equiv u_{\rm rms}D/\nu$ has a mean equal  to 1200, with values up to 3500; one expects that its Reynolds number \emph{relative} to the local fluid motions is of the same order of magnitude, too. 

\begin{figure}[t!]
\centering
\includegraphics[width=\columnwidth]{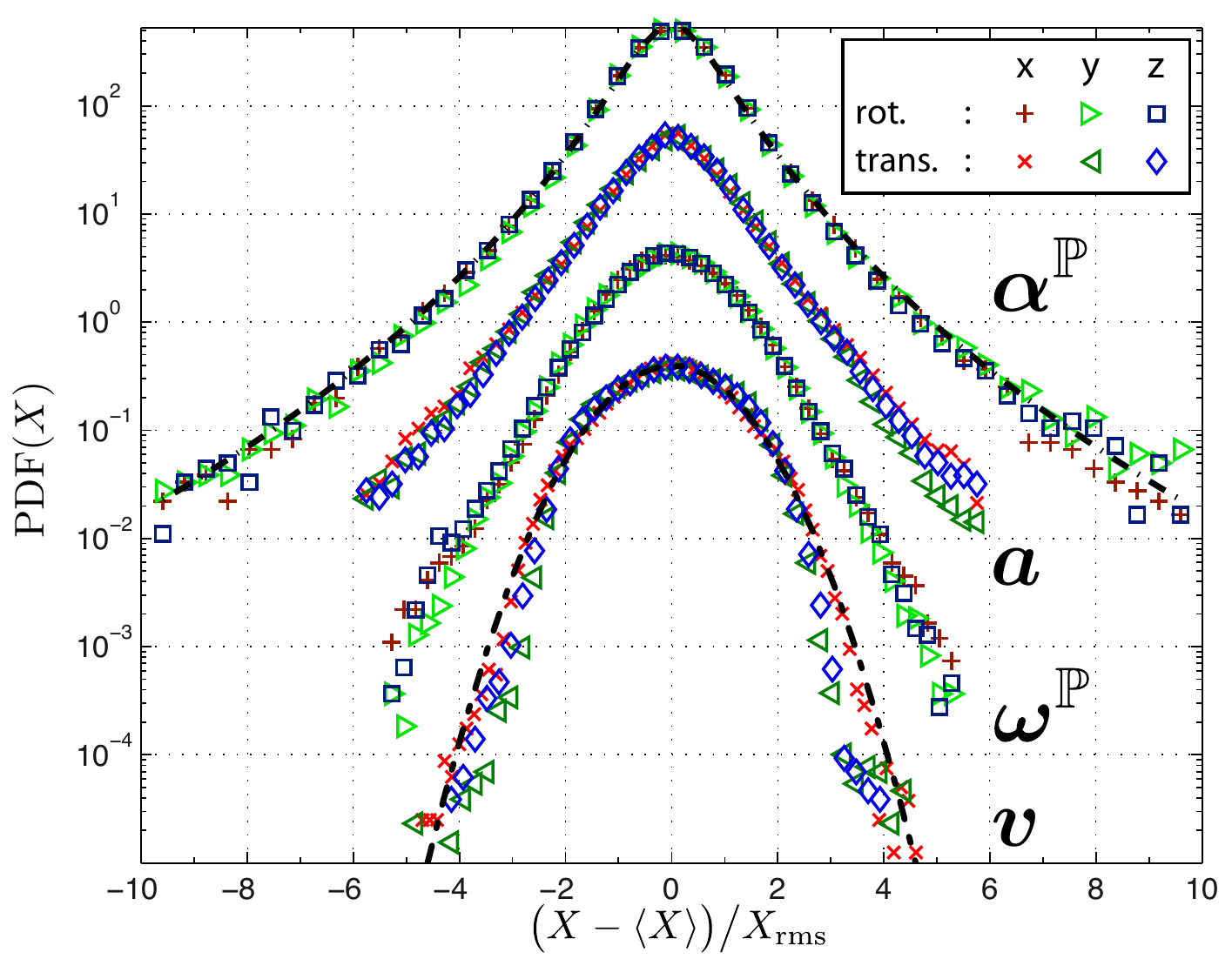}
\caption{PDFs of velocity and acceleration, for the linear and angular motions. The curves have been shifted vertically for clarity.}
\label{pdfs} 
\end{figure}

The particle's acceleration, with zero mean, has an $rms$ amplitude of fluctuations equal to $8.5\,{\rm m/s}^{2}$. Its probability distribution function (PDF) has strongly non-Gaussian fluctuations as shown in \fig{pdfs}.  In terms of velocity increments, the acceleration can be viewed as a velocity change over the shortest time scale, while the velocity increments over long times scales have the same distribution as the velocity itself. Our results thus demonstrate that the PDFs of velocity time increments evolve from a shape with wide tails~\cite{Mordant:2004dk} for very small time lags to a Gaussian shape at long time intervals, 
a phenomenon often called intermittency. This effect is observed here 
for a very large particle, compared to the flow Kolmogorov scale, $\eta$. The strong 
events in the acceleration PDF are, however, less intense than in the case of smaller particles. 
We estimate the normalized fourth moment, the flatness to be 
$F = 7\pm1$  for our large particle ($D/L_{\rm int} \sim 0.6$). 
Remarkably, this value is identical to the one found in~\cite{Qureshi:2008qe}, with a comparable ratio $D/L_{int} \lesssim 1$, and generally with the very slow decay of the flatness with $D$ observed in~\cite{Volk:2010oz,Homann:2010th} for particles much smaller compared to $L_\text{int}$ ($D/L_\text{int} \sim 1/50$). 
\\

We now turn to the angular velocity of the sphere. The three components 
fluctuate around a zero mean value (there is no preferred orientation) and their $rms$ amplitude is 12~rad/s. This value is of the order of the rotation rate of the driving disks, and also corresponds to the rotation that would result from imposing a velocity difference of the order of $u_{\text{rms}}$ across the particle diameter $D$ ($u_{\text{rms}}/D \simeq 30$~rad/s). The PDF of angular velocity components are shown in \fig{pdfs}. The distributions are symmetric, and slightly non-Gaussian (we estimate a flatness factor $F \sim  4$). The $rms$ amplitude of the angular acceleration is about 700~rad/${\rm s}^2$, again of the order of $(u_{\text{rms}}/D)^2$. The statistics of angular acceleration is strongly non-Gaussian (we estimate $F = 7\pm1$). 
Hence, the PDFs of the angular velocity increments becomes broader when
the time-lag $\tau$ decreases from  $\tau \sim T_L$ to $\tau \approx \tau_\eta$: the angular dynamics is intermittent.  \\ 

\begin{figure}[t!]
\centering
\includegraphics[width=0.9\columnwidth]{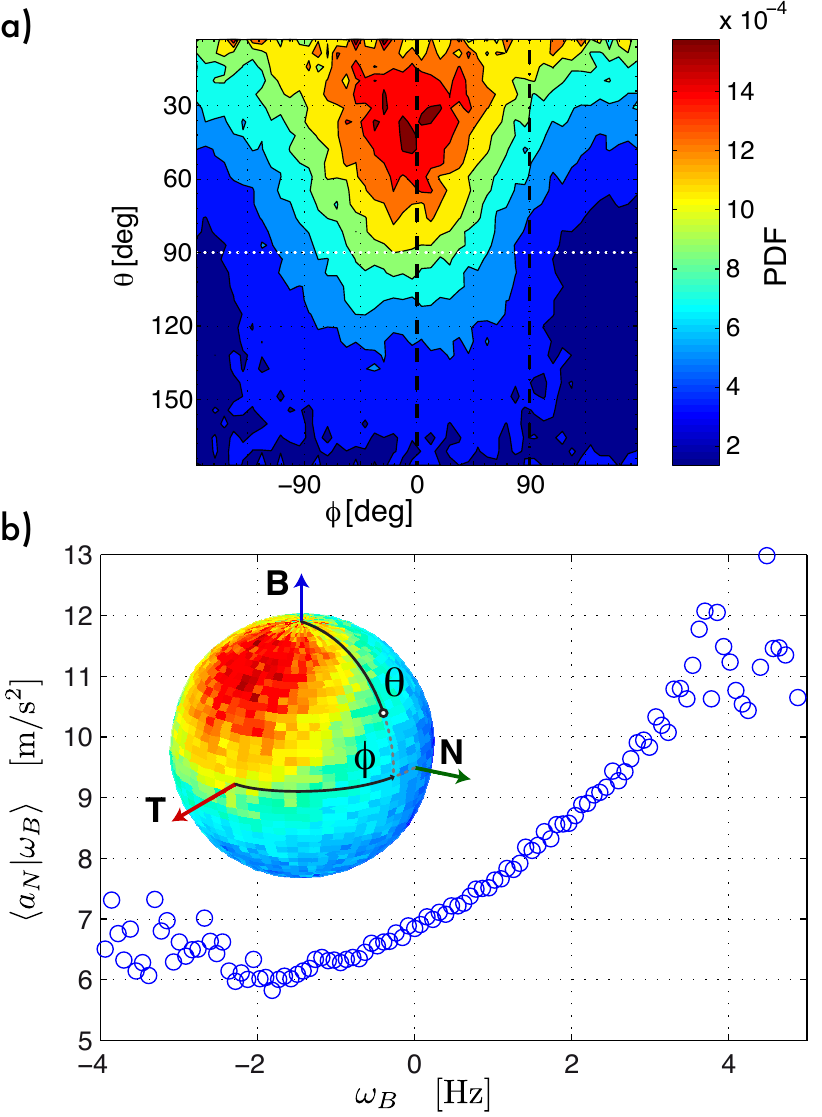}
\caption{(a) alignment of the particle rotation, $\vec\omega\pa$, with respect to the moving Frenet coordinate system; (b) normal acceleration conditioned to the component of angular velocity parallel to the bi-normal Frenet vector.}
\label{coupl_ang_tran} 
\end{figure}

We now address the question of the coupling between the dynamics of rotation 
and translation. \fig{coupl_ang_tran}(a) reveals a strong alignment between the direction of the rotation vector $\vec{\omega\pa}$  and the vectors defining the trajectory  -- the usual Frenet coordinate system ($\vec{T}$,~$\vec{N}$,~$\vec B$), denoting respectively the units vectors along the velocity, the curvature, and the direction perpendicular to the tangent plane of the trajectory. 
We find that $\vec\omega\pa$ is aligned perpendicular to $\vec{N}$, so both the acceleration and $ \vec{\omega\pa} \times\vec{v}$ lie in the ($\vec T, \vec N$)-plane.
While the observation that $\vec\omega\pa \cdot \vec N \approx 0$ is consistent  with a Magnus force, the fairly sharp distribution of the direction 
of $\vec\omega\pa$ on the sphere is unexpected. 
{\fig{coupl_ang_tran}(a) shows a maximum of probability at an angle of the order of $35\degree$ with respect to $\vec{T}$ in the ($\vec{T}, \vec{B}$)-plane. The amplitude of the acceleration (\ie force) along the vector $\vec{N}$, conditioned on the amplitude of $\omega_B \equiv \vec\omega\pa \cdot \vec B$ is shown in \fig{coupl_ang_tran}(b). The averaged normal acceleration, conditioned on $\omega_B$ increases from $6$ to $9 \; {\rm m/s}^2$ (half a standard deviation) when the particle rotation varies in the range  $\pm 12 \; {\rm rad/s}$ (\ie one standard deviation in rotation speed). \\

The main results of this work concern the strong intermittency both for the 
translational and the angular accelerations, and the coupling between the 
rotational and translational degrees of freedom.

The observed intermittency of the translational and rotational accelerations 
points to very intense fluctuations of the force and torque applied to the particle. The interaction between the sphere and the flow involves a pressure and a viscous terms, resulting from the stress tensor: $\tau_{ij} = - p \delta_{ij} + \rho \nu (\partial_i u_j + \partial_j u_i) $. Measurements of the drag force acting on a particle a large Reynolds numbers~\cite{Achenbach:1972ai} demonstrate that pressure is the dominant effect in the force: at the particle's Reynolds number in our experiment,  $Re_P \approx 10^3$, the friction force contributes of the order of $\sim 20 \%$ of the drag; this fraction diminishes at higher Reynolds numbers. 
Pressure is expected to be mostly coherent at the scale of our particle $D \sim L$, so the pressure effects act collectively on the object,  resulting in a force of order $\rho D^2 u_\text{rms}^2$, which yields the correct order of magnitude for the ${rms}$ of the particle acceleration. The strongly non-Gaussian PDF of acceleration fluctuations are more surprising, since the force acting on the particle results from an averaging over a size comparable to the correlation length of the flow.

The effect of friction, although weak compared to pressure, is essential to understand the torque. In fact, the pressure force,  perpendicular to the surface of the sphere, does not contribute to the torque applied at the center of the particle. Thus, the ${rms}$ of the angular acceleration is $\langle \dot{\omega} ^{2} \rangle^{1/2} \sim \langle \Gamma_v^2 \rangle^{1/2}/I$, where $\Gamma_v$ is the viscous torque and $I$ the moment of inertia of the particle. The magnitude of $\Gamma_v$ can be  inferred from the estimated contribution of the friction force to the drag, $F_v \big/F_D  \sim \nu \partial_x u \,\big/ u_\text{rms}^2 \sim u_*^2 \big/u_\text{rms}^2$, where $u_*$ is the velocity characteristic of skin friction velocity, defined by ${u_*}^2 \equiv \nu \partial_x u$. Given the weak value of the moment of inertia tensor, one obtains $\langle \dot{\omega}^2 \rangle^{1/2} \sim 10 (u_*/u_{\rm rms})^2 ( {u_{\rm rms}}/{D})^2$. With $(u_*/u_\text{rms})^2 \sim 0.2$ in the range of Reynolds numbers of the particle motions, this leads to $\langle \dot{\omega}^2 \rangle \sim (u_{\rm rms}^2/D^2)^2$. Which properties of the turbulent flow control the rate of rotation of the particle also remains to be elucidated. In this respect, the naive vision of many small eddies, compared to the size of our big particle, implicit in \fig{flow_sketch}, is likely to be an oversimplification. Small eddies acting on the particle in a spatially incoherent manner would result in a significantly reduced torque acting on the particle. This suggests a much more coherent flow pattern, at least at our particle Reynolds numbers, in fact consistent with the  recent numerical results of \cite{Naso:2010fv}. We also note that while the estimates above provide a qualitative explanation of the observed velocities and accelerations ${rms}$ values, understanding the complete statistics of their fluctuations remains a challenge.

Our evidence of the lift effect, \ie of a strong coupling between rotation and translation, rests on conditioning the acceleration (force) on the angular velocity of the particle, $\vec \omega\pa$. While this is a very sensible choice, the actual torque acting on the particle depends on the local structure of the flow, which remains to be determined, for instance by local Particle Image (or Tracking) Velocimetry around the moving sphere~\cite{Gibert:mi}, or numerically~\cite{Homann:2010th,Lucci:2010bs,Naso:2010fv}. The sharply peaked distribution of $\vec\omega\pa$ in the Frenet basis, see \fig{coupl_ang_tran}(a), is a very surprising result of this work.  In this respect, determining experimentally the variation of the quantities studied here as a function of particle size and Reynolds number should provide  important clues  on the interaction between turbulent flow and the particle. \\

\begin{acknowledgments} This work is part of the International Collaboration for Turbulence Research. We thank Aurore Naso for many fruitful discussions. This work was supported by ANR-07-BLAN-0155, and by PPF `Particules en Turbulence' from the Universit\'e de Lyon.
\end{acknowledgments}
\bibliography{biblioLift}

\end{document}